\documentclass[aps, prd, preprint, nofootinbib, amsfonts, floatfix]{revtex4}

\usepackage{graphicx}
\usepackage{subcaption}
\usepackage{amsmath,amsfonts}
\usepackage{amssymb}
\usepackage{dsfont}
\usepackage{url}
\usepackage{braket}
\usepackage{slashed}
\usepackage{epsfig,color}
\usepackage[dvipsnames]{xcolor}
\usepackage{comment}

\usepackage{tikz}
\usepackage{graphicx}
\usetikzlibrary{positioning}

\newcommand{\be}{\begin{eqnarray}}
\newcommand{\ee}{\end{eqnarray}}

\newcommand\nn{\nonumber}

\newcommand{\mat}{\left ( \begin{array}{cc}}
\newcommand{\emat}{\end{array} \right )}
\newcommand{\vect}{\left ( \begin{array}{c}}
\newcommand{\evect}{\end{array} \right )}

\newcommand{\m}{m_q}

\definecolor{red}{rgb}{1.00, 0.00, 0.00}

\definecolor{blue}{rgb}{0.00, 0.00, 1.00}
\definecolor{green}{rgb}{0.20, 0.6, 0.1}

\definecolor{darkgreen}{rgb}{0.0, 0.4, 0.0}

\long \def \blockcomment #1\endcomment{}

\begin{document}

\title{Quantum Phase Estimation and the Aharonov-Bohm effect}

\author{K. Splittorff}
\affiliation{\hspace{1mm} NNF Quantum Computing Programme, Niels Bohr Institute, University of Copenhagen, Denmark, Blegdamsvej 17, DK-2100, Copenhagen {\O}, Denmark}

\date{\today}
\begin{abstract}
We consider the time evolution of a particle on a ring with a long solenoid through and show that due to the Aharonov-Bohm effect this system naturally makes up a physical implementation of the quantum phase estimation algorithm for a $U(1)$ unitary operator. The implementation of the full quantum phase estimation algorithm with a $U(N)$ unitary operator is realised through the non-abelian Aharonov-Bohm effect. The implementation allows for a more physically intuitive understanding of the algorithm. As an example we use the path integral formulation of the implemented quantum phase estimation algorithm to analyse the classical limit $\hbar\to0$.

\end{abstract}

\maketitle

% INTRODUCTION

\section{Introduction}

The arguably most important quantum algorithm is quantum phase estimation \cite{Kitaev}. It enters as the central building block in quantum algorithms, such as for example Shors algorithm \cite{Shor}, which provide an exponential speed-up as compared to classical algorithms. 
 The implementation of the quantum phase estimation algorithm on a universal quantum computer, is made using a number of qubits and operations thereon, see eg.~the book \cite{NC} or the lecture notes \cite{SA}.

The elegant nature of the quantum phase estimation algorithm suggest an alternative: Could there be a single quantum system with many quantum states which naturally collaborate to implement the quantum phase estimation algorithm? In other words a dedicated quantum computer device that can do quantum phase estimation, but is not a universal quantum computer build out of qubits. Here we point out that there is indeed a physical system which does just that, namely a particle on a ring with a long solenoid through. We show that due to the Aharonov-Bohm effect \cite{AB} the states of this system naturally undergoes a time evolution which makes up a physical implementation of the quantum phase estimation algorithm. The continuum of position states around the ring make up the first register in the quantum phase estimation algorithm\footnote{Note that we here use a single continuous variable rather than a register of qubits. Universal quantum computing over continuous variables have been studied in \cite{LB} and Grovers search algorithm \cite{Grover} has been implemented using continuous variables in \cite{PBL}. For a review of quantum information with continuous variables, see \cite{BvanL}.} and the information about the phase of the eigenvalue of the unitary operator from the second register is induced by the flux through the solenoid. Initially we will restrict ourselves to the case where there is an electromagnetic flux through the ring. This system will implement the quantum phase estimation algorithm for a $U(1)$ unitary operator. To implement the full quantum phase estimation algorithm with a $U(N)$ unitary operator we then extend to the non-abelian Aharanov-Bohm effect \cite{WuYang}.  

Having a physical system which implements a quantum algorithm can help us understand the algorithm better\footnote{One example of this is \cite{Lloyd} where a physical system which realises Grovers algorithm helped clarify the role of entanglement in obtaining quantum advantage. An other example is \cite{CEMM} where a Mach–Zehnder interferometer which solves the Deutsch’s problem \cite{Deutsch} helps highlight the importance of quantum interference.}. Here, we use the implementation by means of the Aharanov-Bohm effect to analyse the classical limit, $\hbar\to0$, of the quantum phase estimation algorithm. To this end we consider the path integral formulation for the implemented quantum phase estimation algorithm and show that, the time it takes to complete the algorithm diverges and the path which minimise the action does {\sl not} dominate in the classical limit $\hbar\to0$. The physical implementation of the quantum phase estimation algorithm by means of the Aharonov-Bohm effect hence does not have a standard classical limit. While perhaps surprising at first, in retrospect this is quite natural, as otherwise the classical system corresponding to the classical limit of the implemented quantum algorithm could potentially {\sl 1)} solve the same task as the quantum algorithm and {\sl 2)} be simulated efficiently on a classical computer.

The paper is organised as follows. First we will recall the steps of the quantum phase estimation algorithm as well as the Aharonov-Bohm effect for a particle on a ring. Then we consider the time evolution of an initially localised particle on the ring and establish the connection to the quantum phase estimation algorithm. The full implementation of the quantum phase estimation algorithm through the non-abelian Aharanov-Bohm effect is then presented. Finally we discuss the Aharonov-Bohm effect in the path integral formalism and the classical limit of the implemented quantum phase estimation algorithm.

\section{Quantum Phase Estimation}

To set the notation we start by briefly recalling the quantum phase estimation (QPE) algorithm \cite{Kitaev}. Given a unitary operator $U$ and an eigenstate $|u\rangle$ of $U$ where\footnote{Note that $\phi_u\in[-\pi,\pi]$ whereas often in the literature a factor of $2\pi$ is factorised such that $\phi_u$ is a number between 0 and 1. Our convention is more natural when we turn to the particle on the ring.}
\begin{eqnarray}
\label{ev-eq}
U|u\rangle = e^{i\phi_u}|u\rangle \ ,
\end{eqnarray}
the aim of the algorithm is to provide an estimate of the phase $\phi_u$. 
The steps of the quantum phase estimation algorithm are listed in FIG.~\ref{Fig:QPE}. Note that the exponent of the phase factor, $e^{i\phi_u j}$, after the controlled unitary operations have been applied, is linear in $j$ and in $\phi_u$.

\begin{figure}[h]
\begin{center}
\begin{tabular}{ l l}
{\bf Operation} & {\bf State}  \\
\hline
 Initial state & $|0\rangle|u\rangle$  \\ 
 QF-transform & $\frac{1}{\sqrt{2^t}}\sum_{j=0}^{2^t-1}|j\rangle|u\rangle$  \\  
 Controlled U's &  $\frac{1}{\sqrt{2^t}}\sum_{j=0}^{2^t-1}e^{i\phi_u j}|j\rangle|u\rangle$  \\
 QF-transform$^{-1}$ & $\frac{1}{2^t}\sum_{j,k=0}^{2^t-1}e^{i(\phi_u-2\pi\frac{k}{2^t}) j}|k\rangle|u\rangle$\\
 Measurement & $|k'\rangle|u\rangle$  \\
 \hline
\end{tabular}
\end{center}
\caption{\label{Fig:QPE} The steps of the quantum phase estimation algorithm \cite{Kitaev}. Note the term in the exponent in the third line which is linear in $\phi_u$ and $j$. The final measurement gives the estimate $2\pi k'/2^t$ for the phase.}
\end{figure}

\section{Intuition}

Our aim is to find a single quantum system which naturally implements the quantum phase estimation algorithm. Before we get into the details, we here give a short (and hopefully) intuitive picture of why we are naturally lead to consider a particle on a ring with a long solenoid through.

We seek a system in which there is a variable which when measured gives us the phase we seek to estimate. This leads us to consider a particle on a ring, where the position along the ring is described by a single angular variable $\phi$. In this setting the quantum Fourier transforms employed in the QPE algorithm transforms between the states with definite value of the angle $\phi$ and the basis of states with definite angular momentum. Now as pointed out above the phases induced by the controlled unitary operations in the QPE algorithm must be linear in $j$ and $\phi_u$. Our aim is to implement these phases through a time evolution of the particle on the ring. Hence we must find a way to induce a Hamiltonian which is linear in the angular momentum quantum number. Now the standard Hamiltonian for the particle on the ring is quadratic in the angular momentum, due to the kinetic term. However, if we introduce a minimally coupled gauge field it will naturally provide the desired linear term in the angular momentum. To get such a minimal coupling to a gauge field into play we introduce a long solenoid through the ring, such that the vector potential associated with the magnetic flux through the solenoid becomes the minimally coupled gauge field. Moreover the new term which is linear in angular momentum will also automatically be linear in the magnetic flux through the solenoid. Hence the information about the phase of the eigenvalue of the unitary operator can be induced through the magnetic flux. However, in the Hamilton operator we are still left with the original term quadratic in the angular momenta, and there are no analogous terms in the QPE algorithm. Fortunately, it is possible to nullify the effect of this quadratic term by picking the evolution time for the system carefully. The Aharanov-Bohm effect will in this way implement the QPE algorithm for a $U(1)$ unitary operator. To implement the QPE algorithm with a $U(N)$ unitary operator we extend to the non-abelian Aharanov-Bohm effect where the particle is minimally coupled to a $U(N)$ gauge field.

Hopefully, this intuitive picture can serve as a guide through the subsequent sections. 

\section{Particle on a ring}

For a start let us coinsider a particle on a ring of radius $r$, described by the Hamiltonian 
\begin{eqnarray}
H^0 = \frac{L_z^2}{2\m r^2}=-\frac{\hbar^2}{2\m r^2}\frac{\partial^2}{\partial\phi^2} \ .
\end{eqnarray}
Here $\phi\in[-\pi,\pi]$ is the angular coordinate of the particle and
\begin{eqnarray}
L_z=-i\hbar\frac{\partial}{\partial\phi}
\end{eqnarray}
is the $z$-component of the angular momentum of the particle about the center of the ring. (Our coordinate system is such that the ring is in the $(x,y)$-plane with the center at the origin.) We use $\m$ to denote the mass of the particle. 

The normalised eigenfunctions of $H^0$ are
\begin{eqnarray}
\label{psim}
\psi_m(\phi)=\frac{1}{\sqrt{2\pi}}e^{im\phi}
\end{eqnarray}
with corresponding eigenenergies
 \begin{eqnarray}
E^0_m=\frac{\hbar^2}{2\m r^2}m^2 \ ,
\end{eqnarray}
where $m=\ldots,-1,0,1,\ldots$.
\bigskip

\subsection{Time evolution and quantum mechanical return time} 

Next let's consider the time evolution of the state of the particle on the ring. If at time $t=0$ the particle is in the state 
\begin{eqnarray}
\label{Psi_t0}
\Psi(\phi,t=0) = \sum_{m=-\infty}^\infty c_m \psi_m(\phi)
\end{eqnarray}
then the state at time $t$  is obtained by tagging on the wiggle-factors, $e^{-iE^0_mt/\hbar}$, that is
\begin{eqnarray}
\Psi(\phi,t) = \sum_{m=-\infty}^\infty c_m e^{-iE^0_mt/\hbar}\psi_m(\phi) \ .
\end{eqnarray}
Note that since $E^0_m=\frac{\hbar^2}{2\m r^2}m^2$ and $m$ is an integer, all the wiggle-factors will equal 1 at the quantum mechanical return time 
\begin{eqnarray}
\label{t-return}
t_R =  \frac{4\pi \m r^2}{\hbar} \ .
\end{eqnarray}
The state of the particle will therefore return to its initial state at $t = t_R$ for {\sl any} initial state of the particle on the ring.

\bigskip

\subsection{Time evolution of an initially localised particle.}

For example, if the particle is fully localised at $\phi=0$ for time $t=0$  
\begin{eqnarray}
\Psi(\phi,t=0) = \delta(\phi) \ ,
\end{eqnarray}
then with time the state will become delocalised. However, at time $t=t_R$ the state of the system will again be localised at $\phi=0$, that is $\Psi(\phi,t_R) = \Psi(\phi,t=0)=\delta(\phi)$.

In order to see this explicitly, we represent the $\delta$-function as
\begin{eqnarray}
\label{delta}
\Psi(\phi,t=0) = \delta(\phi)=\lim_{l\to\infty}\frac{1}{\sqrt{2\pi(2l+1)}}\sum_{m=-l}^l e^{im \phi} \ ,
\end{eqnarray}
and note that this can be re-written as
\begin{eqnarray}
\Psi(\phi,t=0) = \delta(\phi)=\lim_{l\to\infty}\frac{1}{\sqrt{2l+1}}\sum_{m=-l}^l \psi_m(\phi)\ .
\end{eqnarray}
The initial state of the system is now expressed in terms of the stationary states of $H$ and the time evolution of the state is obtained by tagging on the   wiggle-factors, which for $t = t_R$ all equal 1. So at $t=t_R$ the particle initially localised at $\phi=0$ will again be localised at $\phi=0$.

\section{The Aharonov-Bohm effect for a particle on a ring}
\label{sec:AB}

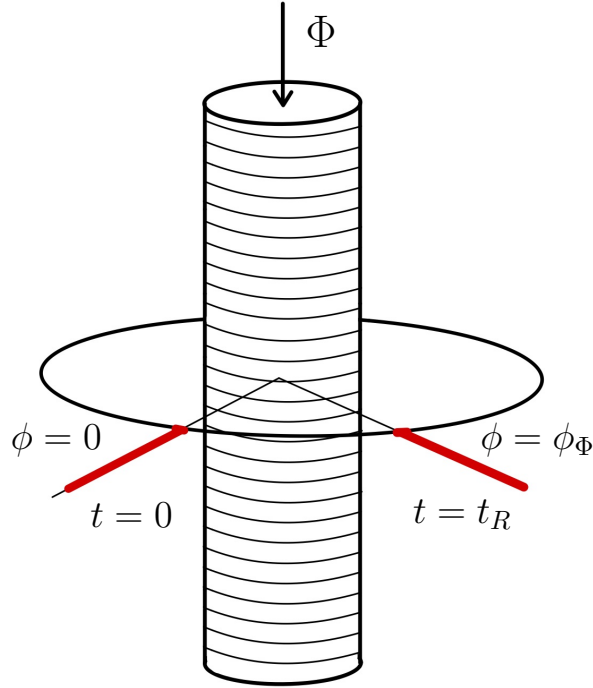
\begin{figure}[htb]
   \vspace{-1.5cm}
\begin{minipage}{\textwidth}
\begin{tikzpicture}
  \node (img)  {\includegraphics[scale=0.2]{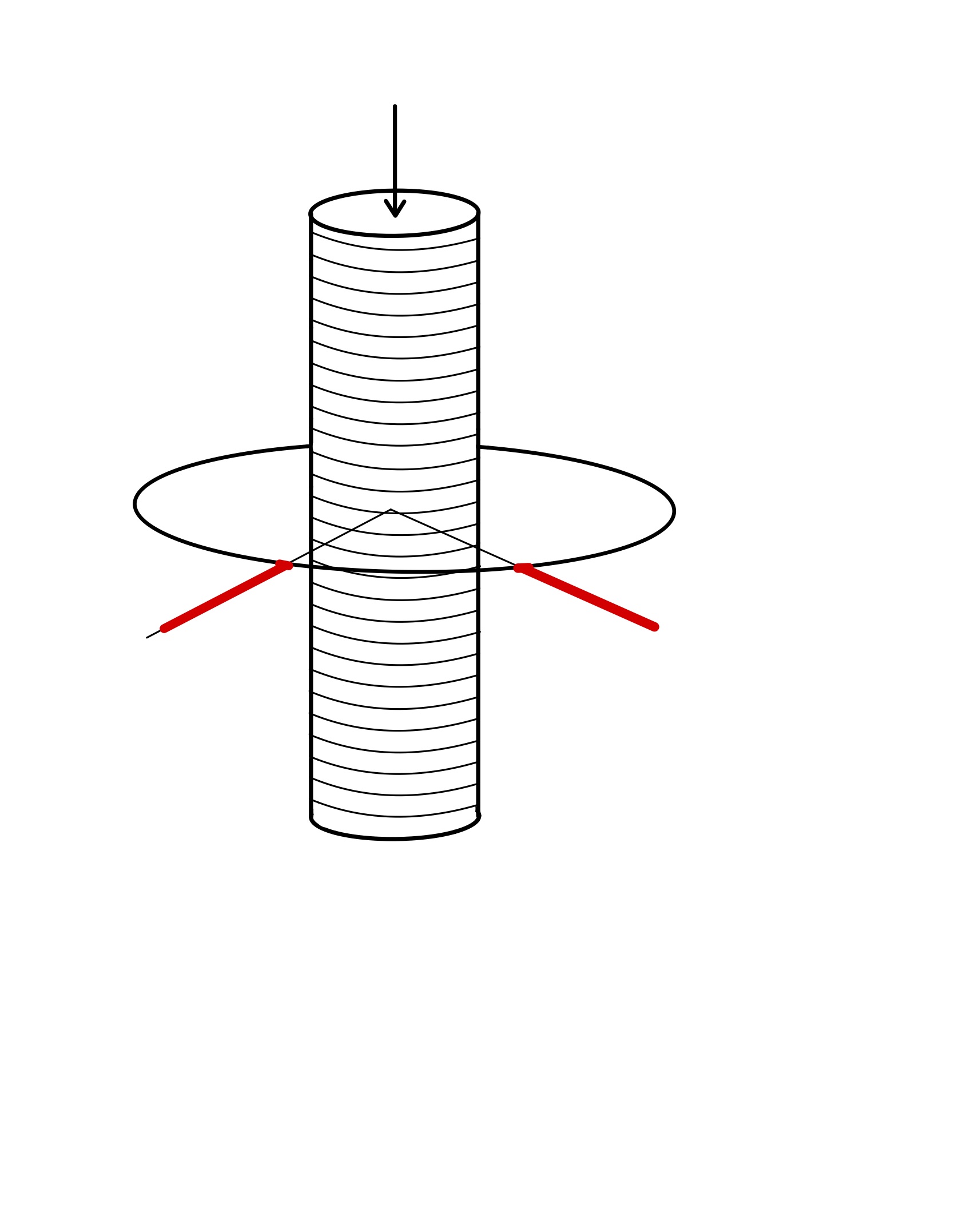}};
  \node[below=of img, node distance=0cm, yshift=15cm,xshift=-0.5cm,font=\color{black}] {\Large $\Phi$};
  \node[left=of img, node distance=0cm, rotate=0, anchor=center,yshift=0.5cm,xshift=3.0cm,font=\color{black}] {\large $\phi=0$};
    \node[left=of img, node distance=0cm, rotate=0, anchor=center,yshift=-0.5cm,xshift=4.0cm,font=\color{black}] {\large $t=0$};
  \node[left=of img, node distance=0cm, rotate=0, anchor=center,yshift=0.5cm,xshift=9.4cm,font=\color{black}] {\large $\phi=\phi_\Phi$};
    \node[left=of img, node distance=0cm, rotate=0, anchor=center,yshift=-0.5cm,xshift=8.4cm,font=\color{black}] {\large $t=t_R$};

 \end{tikzpicture}
\end{minipage}%
    \vspace{-4.5cm}
     \caption{\label{fig:AB-QPE} A particle on a ring with a long solenoid through. At time $t=0$ the particle is localised at $\phi=0$ and at $t=t_R$ it is localised at $\phi_\Phi=-2\frac{q\Phi}{\hbar}$, where $\Phi$ is the flux through the solenoid. The localised wave function is illustrated in red, see FIG.~\ref{fig:AB-time-development} for a plot.}
\end{figure}

We now place a solenoid through the ring such that its symmetry axis is identical to the $z$-axis and the radius of the solenoid is smaller than that of the ring, see FIG.~\ref{fig:AB-QPE}. In the limit where the length of the solenoid is much greater than the radius of the ring the magnetic field at the ring is zero, but newer the less the state of the particle on the ring will affected by the solenoid, through the Aharanov-Bohm effect \cite{AB}. The Aharanov-Bohm effect is mediated by the vector potential $A$ which will be non-zero outside the solenoid.

For a static flux $\Phi$ through the solenoid the Hamiltonian for the particle on the ring becomes
\begin{eqnarray}
H =\frac{1}{2\m}\left(-i\frac{\hbar}{r}\frac{\partial}{\partial\phi}-q\frac{\Phi}{2\pi r}\right)^2 \ ,
\end{eqnarray}
where $q$ is the charge of the particle. 
Note that the shift of the derivative is realised through the minimal coupling of the vector potential of magnitude $\Phi/(2\pi r)$, see e.g.~\cite{GS}.

Because the rotational symmetry around the $z$-axis remains, the eigenfunctions (\ref{psim}) of $H$ are unaffected by the presence of the solenoid. However, the corresponding eigenenergies now become
 \begin{eqnarray}
 \label{EofPhi}
E_m=\frac{\hbar^2}{2\m r^2}\left(m-\frac{q\Phi}{2\pi\hbar}\right)^2 \ .
\end{eqnarray}
\bigskip

\subsection{Time evolution and re-appearance at shifted angle} 

As a first step towards establishing the connection to the quantum phase estimation algorithm let us now consider the time evolution of the state of the particle on the ring in the presence of a flux through the solenoid.

Because the wiggle-factors are determined by the energies and these depend on $\Phi$ cf.~(\ref{EofPhi}) the solenoid affects the time evolution of the state of the particle on the ring. At non-zero $\Phi$ the eigenenergies, $E_m$, no longer have the simple relationship of some number times an integer which depends on $m$, hence the initial state of the system will no longer automatically reappear at a later time. However, if we evolve the system for the quantum mechanical return time (\ref{t-return}) obtained for $\Phi=0$,  the state of the system returns to the exact initial form {\sl except} that it is shifted by an angle $\phi_\Phi$ determined by the flux $\Phi$. 

To determine the shift $\phi_\Phi$ we again start the particle in an arbitrary normalised state (\ref{Psi_t0}) and tag on the wiggle-factors, $e^{-iE_mt/\hbar}$, 
\begin{eqnarray}
\label{Psi-with-Phi}
\Psi_\Phi(\phi,t) 
               & = & e^{-i\frac{\hbar}{2\m r^2}(\frac{q\Phi}{2\pi\hbar})^2t}\sum_{m=-\infty}^\infty c_m e^{-i\frac{\hbar}{2\m r^2}m^2t}e^{i\frac{\hbar}{2\m r^2}\frac{q\Phi}{\pi\hbar}mt}\psi_m(\phi) \ . 
\end{eqnarray}
Note in particular the exponent which is proportional to the flux $\Phi$ and $m$. This is exactly the desired term which will allow us to implement the quantum phase estimation algorithm by means of the Aharanov-Bohm effect. 

Inserting the eigenstates $\psi_m(\phi)$ of (\ref{psim}) we get
\begin{eqnarray}
\label{Psi-with-Phi-v2}
\Psi_\Phi(\phi,t)  
                & = & \frac{1}{\sqrt{2\pi}}e^{-i\frac{\hbar}{2\m r^2}(\frac{q\Phi}{2\pi\hbar})^2t}\sum_{m=-\infty}^\infty c_m e^{-i\frac{\hbar}{2\m r^2}m^2t}e^{im(\phi + \frac{\hbar}{2\m r^2}\frac{q\Phi}{\pi\hbar}t)}  \\
                & = & e^{-i\frac{\hbar}{2\m r^2}(\frac{q\Phi}{2\pi\hbar})^2t}  \Psi_{\Phi=0}(\phi+\frac{\hbar}{2\m r^2}\frac{q\Phi}{\pi\hbar}t,t) \ .\nonumber
\end{eqnarray}
So up to an overall phase factor, which does not affect the probability distribution, we have 
 \begin{eqnarray}
\Psi_\Phi(\phi,t) & = &  \Psi_{\Phi=0}(\phi+\omega_\Phi t,t) \ ,\nonumber
\end{eqnarray}
 where we introduced the angular velocity 
 \begin{eqnarray}
 \label{om_Phi}
\omega_\Phi \equiv \frac{\hbar}{2\m r^2}\frac{q\Phi}{\pi\hbar} \ .
 \end{eqnarray}
In words, the wave function of the particle on the ring evolves exactly the same way as it did for zero flux through the solenoid, except that it now in addition rotates with an angular velocity which depends linearly on the flux. So just as for zero flux the initial state of the system will reappear at time $t=t_R$ only now it is shifted around the ring by 
  \begin{eqnarray}
 \phi_\Phi\equiv-\omega_\Phi t_R=-2\frac{q\Phi}{\hbar} .  
 \label{phi_Phi}
  \end{eqnarray}

\begin{figure}[h!]
\begin{center}
\begin{tabular}{ l l l }
{\bf Operation} & {\bf State} & {\bf Physical picture} \\
\hline 
 Initial state & $\delta(\phi)|u\rangle$ & localised at $\phi=0$  \\ 
 Express in $L_z$-basis & $\lim_{l\to\infty}\frac{1}{\sqrt{2l+1}}\sum_{m=-l}^l e^{im\phi}|u\rangle$ & delocalised in $L_z$-basis \\  
Evolve for time $t_R$ &  $\lim_{l\to\infty}\frac{1}{\sqrt{2l+1}}\sum_{m=-l}^l e^{i(\phi - \phi_\Phi)m}|u\rangle$ & delocalised in $L_z$ but with phase input \\
 Express in $\phi$ basis & $\delta(\phi-\phi_\Phi)|u\rangle$ & localised at $\phi=\phi_\Phi$ \\
 Measurement of $\phi$ & $\delta(\phi-\phi_\Phi)|u\rangle$ & result is $\phi_\Phi$ \\
 \hline
\end{tabular}
\end{center}
\caption{\label{tab:AB} The steps in the evolution of an initially localised particle on a ring between time $t=0$ and $t=t_R$ due to the Aharonov-Bohm effect. Note the exponent in the third line which is linear in $\phi_\Phi$ and $m$. Compare to the steps of the QPE algorithm, FIG.~\ref{Fig:QPE}.}
\end{figure}

\section{The Aharonov-Bohm effect and the QPE algorithm}
\label{Sec:AB-phase}

We are now in a position to make the connection between the Aharonov-Bohm effect and the quantum phase estimation algorithm, all we need is to consider an initially localised particle on the ring, $\Psi(\phi,t=0) = \delta(\phi)$. 
In this case we have at time $t_R$ that the particle is fully localised at $\phi_\Phi=-2\frac{q\Phi}{\hbar}$  
\begin{eqnarray}
\Psi(\phi,t=t_R) = \delta(\phi+2\frac{q\Phi}{\hbar}) \ .
\end{eqnarray}
A measurement of the particles position (as given by the angular coordinate) at time $t=t_R$ hence will give $\phi=-2\frac{q\Phi}{\hbar}$. 

The steps are summarised in FIG.~\ref{tab:AB}. Note that they match those of the QPE algorithm,  compare FIG.~\ref{Fig:QPE}. 
\bigskip

\begin{figure}[h!]
 { \centering
  \begin{subfigure}[b]{0.3\linewidth}
    \includegraphics[width=\linewidth]{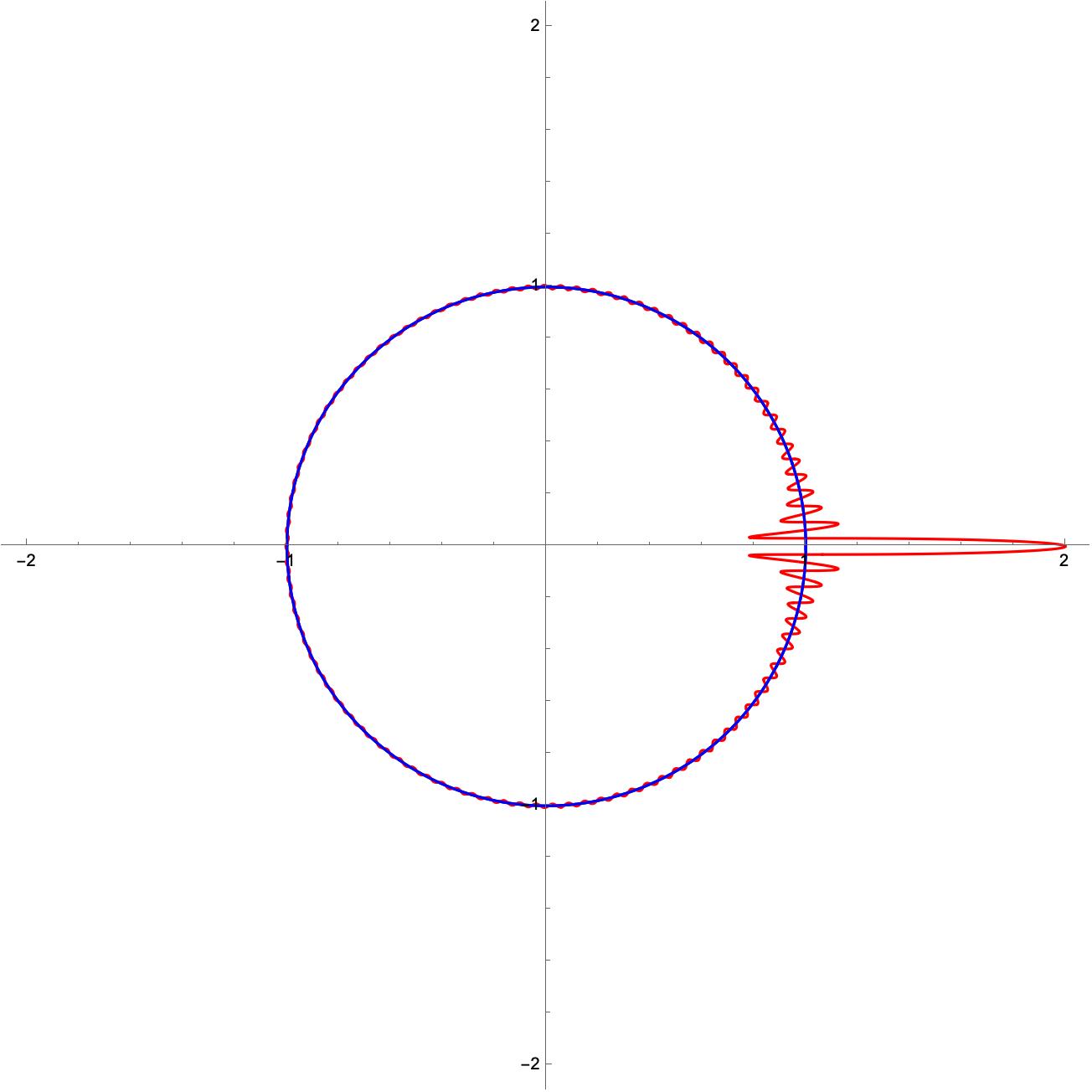}
    \caption{t=0}
  \end{subfigure}
  \begin{subfigure}[b]{0.3\linewidth}
    \includegraphics[width=\linewidth]{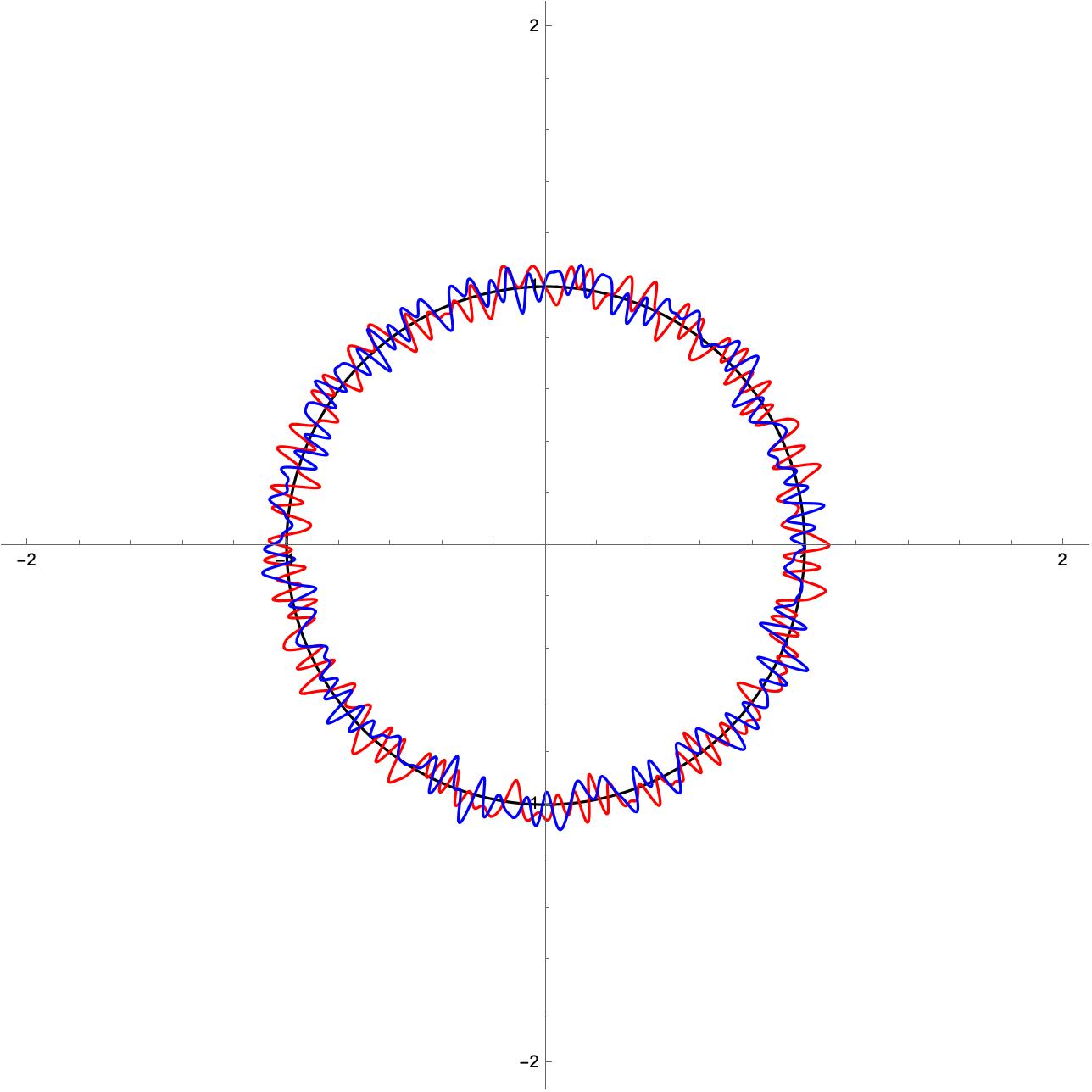}
    \caption{$t=t_R/\sqrt{2}$}
  \end{subfigure}
   \begin{subfigure}[b]{0.3\linewidth}
    \includegraphics[width=\linewidth]{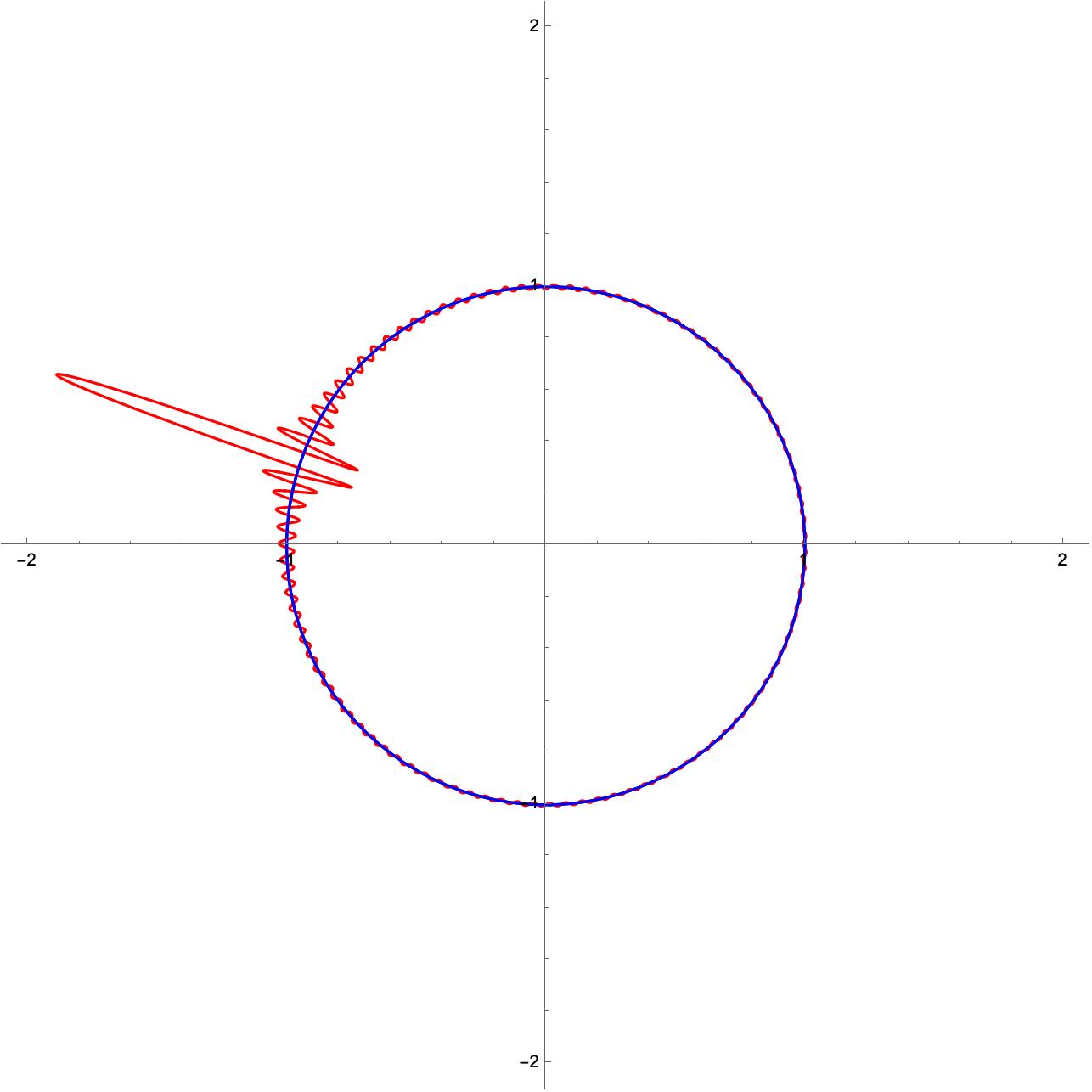}
    \caption{$t=t_R$}
  \end{subfigure}}
  \caption{\label{fig:AB}The time evolution of the wave function of a particle on the ring with a long solenoid through. (a) Initially the particle is localised at $\phi=0$. (b) It then delocalises and finally (c) at $t=t_R$ it is localised at $\phi=-2\frac{q\Phi}{\hbar}$, where $\Phi$ is the flux through the solenoid. Red: real part of the wave function. Blue: imaginary part. For the illustration we have chosen $l=100$ in (\ref{delta}).}
  \label{fig:AB-time-development}
\end{figure}

\noindent
The corresponding transcription table for the phase is 
\be
\phi_u = \phi_\Phi=-2\frac{q\Phi}{\hbar} \ .
\ee
 For an illustration of the time development of the wave function of the particle on the ring see  Fig.~\ref{fig:AB-time-development}.

To summarise this far: Because the Aharonov-Bohm effect induce a term linear in respectively $m$ and $\phi_\Phi$, and because the effect of the $m^2$-term is nullified at $t=t_R$, the state of the particle on the ring at $t=t_R$ matches that obtained by the QPE algorithm, compare FIG.~\ref{Fig:QPE} and FIG.~\ref{tab:AB}. In the section below we to extend to a non-abelian flux. This will allow us to fully implement the second register and the unitary operator $U$ of (\ref{ev-eq}). Before we turn to this let's first make one cross check of the arguments this far which in addition will be very useful when we extend to the non-abelian Aharanov-Bohm effect.

\subsection{Aharanov-Bohm phase directly}

As stressed in \cite{WuYang} the Aharanov-Bohm effect is due to the phase factor 
\begin{equation}
\label{AB-phase}
\exp\left(i\frac{q}{\hbar}\oint A \cdot dx)\right) \ .
\end{equation}
Here $A$ is the vector potential and the integral is around a closed loop, in our case around the ring. We will now show that this is completely consistent with our analysis this far.

First recall that after the quantum Fourier transform the state of the initially localised particle is expressed in terms of states, $\psi_m(\phi)$, with well defined angular momentum $\hbar m$, see (\ref{delta}). Second, if we equate $\hbar m=m_qr^2\omega$ we have $\omega t_R=2\pi 2m$, ie.~the state $\psi_m(\phi)$ winds around the ring $2m$ times in time $t_R$. Now, as the vector potential, $A$, at the ring has a constant magnitude, $\Phi/2\pi r$, and points along the ring, the integral in (\ref{AB-phase}) gives $2\Phi m$ and we get the Aharanov-Bohm phase factor for the state $\psi_m(\phi)$
\begin{equation}
\exp\left(i2\frac{q}{\hbar}\Phi m\right) \ .
\end{equation}
This phase factor is fully consistent with (\ref{Psi-with-Phi-v2}) at time $t_R$ and thus we reproduce the results of section \ref{Sec:AB-phase}. 

We will use the formulation in terms of the Aharanov-Bohm phase factor as we now extend to the non-abelian Aharanov-Bohm effect and the full QPE algorithm.

\section{The non-abelian Aharanov-Bohm effect}

This far we have considered the flux through the solenoid as an electromagnetic source for the Aharonov-Bohm effect. As we now show the full quantum phase estimation algorithm is implemented when we generalise to a non-abelian flux corresponding to a $U(N)$ gauge field, that is if we consider the non-abelian Aharonov-Bohm effect \cite{WuYang} for the particle on the ring\footnote{Note, that our analysis this far corresponds to the case $U(1)$ and that the particle on the ring in this setting acts as a quantum sensor of the magnetic flux which works by means the quantum phase estimation algorithm.}.

In the non-abelian Aharonov-Bohm effect \cite{WuYang} the vector potential, $A$, is replaced by a vector field, $A^kX_k$, from a non-abelian gauge group which in our case is taken to be $U(N)$. Here $A^k$ for $k=1,\ldots,N^2$ parameterise the gauge fields, $X_k$ are the generators of $U(N)$ and summation on $k$ is implied.  The corresponding non-abelian Aharanov-Bohm phase factor is \cite{WuYang}
\begin{equation}
\label{AB-non-a-phase}
U_{AB} = \exp\left(i\frac{1}{\hbar}\oint A^kX_k \cdot dx\right) \ .
\end{equation}
As the generators $X_k$ are Hermitian matrices the non-abelian Aharanov-Bohm phase factor it self is a unitary $N$ by $N$ matrix and the state, $\psi(\phi)$, of the particle on the ring now carries a corresponding index\footnote{For the gauge group $SU(3)$ we can consider the strong interactions ie.~quantum chromodynamics, where $A^k$ can be thought of as the gluons and the particle on the ring as a quark which carries a color index.} (we can write eg.~$X^{ab}\psi^b(\phi)$ where summation over $b$ is implied). The degree of freedom indicated by the matrix index will make up the second register in the QPE algorithm.

Recall that in the QPE algorithm the state $|u\rangle$ of the second register is picked as an eigenstate of $U$ such that the controlled application of $U$ to the state
\begin{eqnarray}
 |j\rangle|u\rangle & cU \atop \longrightarrow & e^{i\phi_u j}|j\rangle|u\rangle \ .
 \end{eqnarray}
Likewise we pick $\psi_m^a(\phi)$ in the space indicated by the index $a$, such that the non-abelian Aharanov-Bohm phase factor $U_{AB}$ for $\psi_m^a(\phi)$ is diagonal
 \begin{eqnarray}
 \label{non-a-AB-phase-m}
\psi_m^a(\phi) & U_{AB} \atop \longrightarrow & e^{i\phi_{AB}m}\psi_m^a(\phi) \ .
  \end{eqnarray}
Note that the crucial factor of $m$ in the exponent comes automatically: $\psi_m^a(\phi)$ completes $m$ times as many windings around the ring in time $t_R$ as the component with $m=1$ and therefore the loop integral in the non-abelian Aharanov-Bohm phase factor is completed a factor $m$ times more. As we have seen above for the electromagnetic $U(1)$ case the phase factor in (\ref{non-a-AB-phase-m}) is exactly what we need in order to transfer the knowledge about the eigenvalue from the second register to the first. 
This implies that the non-abelian Aharanov-Bohm effect for the particle on the ring will make up a physical implementation of the QPE algorithm for the unitary operator $U_{AB}$ corresponding to the non-abelian Aharanov-Bohm phase factor for $\psi^a_{m=1}$, with a flux determined by the gauge fields $A^k$. As we are free to pick the gauge fields, we conclude that the non-abelian Aharonov-Bohm effect for the particle on the ring makes up a physical implementation of the full QPE algorithm.

\bigskip

The implementation of the QPE algorithm by a physical system gives a more physically intuitive framework in which we may analyse the QPE algorithm\footnote{The technological realisation of the quantum phase estimation algorithm by means of the non-abelian Aharonov-Bohm effect could of course be extremely challenging. Even if manage to set up a non-abelian flux and even if schemes to perform error correction for continuous quantum variables do exist \cite{Braunstein,LS,DKP}, the resolution of the detector \cite{PBL} and the resources needed to perform the final measurement of the phase must be considered \cite{Lloyd}.}. As a first example we now consider the classical limit of the implemented QPE algorithm.

\section{The classical limit of implemented QPE algorithm}

Our aim is to examine the classical limit where $\hbar\to0$ of the implemented QPE algorithm, and hence 
it is natural to consider the system in the path integral formalism\footnote{For discussions of quantum computing, computational paths and path integrals, see e.g.~\cite{BV,DHHMNO,PKS}.} \cite{feynman}. In this description one identifies the path for which the action is minimal (extreme) and the standard scenario is that this path dominates the path integral in the classical limit. The path which minimises the action, in this case, becomes the classical path. Suppose that this standard scenario holds true for our single quantum system which implements the QPE algorithm. Then perhaps one could maintain the quantum advantage of the QPE algorithm by using only the classical path and get a remarkable breakthrough in classical computing. Alternatively something (almost) equally exiting must happen in our single quantum system, namely that the path which minimises the action does not dominate the path integral in the classical limit. As we show below it is this latter option which is realised. 

\subsection{Path integral approach}  

Path integrals offer a way to analyse and compute the quantum mechanical transition amplitude between the state of the system at two different times \cite{feynman}. To be more precise, if we know that the particle is in a state with a certain well defined position at time $t=0$, the path integral allow us to compute the quantum mechanical amplitude to detect the particle at a given position at a later time $t$.   The fundamental approach is to subdivide the time interval $[0,t]$ into many smaller intervals and in each of these to insert complete sets of respectively position and momentum states into the transition amplitude. In each of these tiny time intervals the time evolution operator acts for a short duration. This leads to the path integral in phase space. Note that this is a path integral over both space and momenta and that these variables are not linked by the equations of motion. Only if the action is quadratic in momentum then the momentum space integral can be performed by a Gaussian integration, resulting in the Feynman path integral representation in configuration space, see eg.~the lecture notes \cite{AP}.

\subsection{The path integral for the Aharonov-Bohm implemented QPE algorithm}

To derive the path integral formulation for the QPE algorithm as implemented by the particle on the ring, we follow the general approach outlined above and split the interval $[0,t_R]$ into $N$ time steps of length $\delta t= t_R/N$. Next we insert a unity in the form of projections on all of angular momentum space, $1=\sum |m_j\rangle\langle m_j|$, before the time interval, and angular space, $1=\int d\phi_l \, |\phi_l\rangle\langle\phi_l|$, after the time interval\footnote{As the second register is unaltered throughout the QPE algorithm we suppress the state $|u\rangle$ below.}
\be
 \langle \phi |e^{-iHt_R/\hbar}|0\rangle 
 & = & \sum_{m_1\ldots m_{N}}\int  d\phi_1d\phi_2 \ldots d\phi_{N} \\
&& \times \langle \phi |\ldots|\phi_2\rangle\langle \phi_2|e^{-iH\delta t/\hbar}|m_2\rangle\langle m_2|\phi_1\rangle\langle \phi_1|e^{-iH\delta t/\hbar}|m_1\rangle\langle m_1|0\rangle \nn 
\ee
where $t_R=N\cdot \delta t$. The states $|m\rangle$ are eigenstates of $H$ with eigenenergies given by (\ref{EofPhi}) and $\langle m|\phi\rangle=\frac{1}{\sqrt{2\pi}}e^{-i \phi m}$. Inserting this we get 
\be
 \langle \phi |e^{-iHt_R/\hbar}|0\rangle & = & \sum_{m_1\ldots m_{N}}\int  d\phi_1d\phi_2 \ldots d\phi_{N} \\
& & \times \ldots e^{i \phi_2 \, m_2} e^{-i\frac{\hbar}{2\m r^2}\left(m_2-\frac{q\Phi}{2\pi\hbar}\right)^2\delta t}e^{-i \phi_1 \, m_2} e^{i \phi_1 m_1}  e^{-i\frac{\hbar}{2\m r^2}\left(m_1-\frac{q\Phi}{2\pi\hbar}\right)^2\delta t} e^{-i 0 m_1} \ . \nn 
\ee
This is the path integral in phase space for the QPE algorithm as implemented by the particle on the ring in the presence of the solenoid. To get the path integral in configuration space we now do the sum over $m_j$ by Poisson summation (for simplicity we do not keep track of the $\phi_j$-independent pre-exponential factors) 
 \be
&&  \sum_{m_j=-\infty}^\infty \ e^{-i\frac{\hbar}{2\m r^2}\left(m_j-\frac{q\Phi}{2\pi\hbar}\right)^2\delta t}e^{i (\phi_{j}-\phi_{j-1}) \, m_j} \\
&& \simeq  
\sum_{n=-\infty}^\infty e^{i\frac{1}{2}\frac{\m r^2}{\hbar} \left(\phi_{j}-\phi_{j-1}+2\pi n\right)^2\frac{1}{\delta t}}e^{i (\phi_{j}-\phi_{j-1}+2\pi n)\frac{q\Phi}{2\pi\hbar}} \ .\nn
\ee
In the limit $\delta t\to 0$ the $n=0$ term dominates the sum. The path integral in configuration space therefore becomes
\be
\label{expiA}
 \langle \phi |e^{-iHt_R/\hbar}|0\rangle 
& \simeq & \int d\phi_1d\phi_2 \ldots \ \prod_{j=1}^{N}e^{i\frac{1}{2}\frac{\m r^2}{\hbar} \left(\phi_{j}-\phi_{j-1}\right)^2\frac{1}{\delta t}}e^{i (\phi_{j}-\phi_{j-1})\frac{q\Phi}{2\pi\hbar}} \\
& = &  \int d\phi_1d\phi_2 \ldots \  e^{i \sum_{j=1}^{N}\frac{1}{2}\frac{\m r^2}{\hbar} \left(\phi_{j}-\phi_{j-1}\right)^2\frac{1}{\delta t}+(\phi_{j}-\phi_{j-1})\frac{q\Phi}{2\pi\hbar}} \nn \\
& = &  \int d\phi_1d\phi_2 \ldots \  e^{i A(\{\phi_j\})/\hbar } \ , \nn
\ee
where $\phi_0=0$ and $\phi_{N}=\phi$ and we introduced the action in configuration space
\be
\label{A-config-space}
A(\{\phi_j\})  =\hbar\sum_{j=1}^{N}\frac{1}{2}\frac{\m r^2}{\hbar} \left(\phi_{j}-\phi_{j-1}\right)^2\frac{1}{\delta t}  + (\phi_{j}-\phi_{j-1})\frac{q\Phi}{2\pi\hbar} \ .
\ee
The path which minimise the action has a constant step size, ie.~$\phi_{j}-\phi_{j-1}$ is independent of $j$,  
\be
\label{k_cl}
\phi_{j}-\phi_{j-1} =-\frac{q\Phi}{2\pi\hbar} \frac{\hbar}{\m r^2} \delta t \ .
\ee
Note that the slope
\be
\label{omega_from_pathint}
\frac{\phi_j-\phi_{j-1}}{\delta t}=-\frac{q\Phi}{2\pi} \frac{1}{\m r^2} \ ,
\ee
can be thought of as an angular velocity and matches the result of (\ref{om_Phi}).
\bigskip

To convince ourselves that the path given by (\ref{k_cl}) indeed minimises the action let us introduce a deviation from this path 
\be
\label{deviation}
\phi_{j}-\phi_{j-1} = - \frac{q\Phi}{2\pi\hbar} \frac{\hbar}{\m r^2} \delta t + \epsilon_j \ .
\ee
Note that $\epsilon_j$ are increments away from the path which minimise the action. At step $j'$ the deviation from this minimising path hence will be $\sum_{j=1}^{j'}\epsilon_j$ and to ensure that the path ends in $\phi_N=\phi$ we have $\sum_{j=1}^{N}\epsilon_j=0$.

As a check that the linear terms in $\epsilon$ of the action vanish we insert (\ref{deviation}) in (\ref{A-config-space}) 
\be
A(\{k_j\}) 
& = & \hbar\sum_{j=1}^{N}\frac{1}{2}\frac{\m r^2}{\hbar} \left( -\frac{q\Phi}{2\pi\hbar} \frac{\hbar}{\m r^2} \delta t + \epsilon_j\right)^2\frac{1}{\delta t}  - \frac{q\Phi}{2\pi\hbar} \frac{\hbar}{\m r^2} \delta t + \epsilon_j\frac{q\Phi}{2\pi\hbar}   \\
& = & \hbar\sum_{j=1}^{N}\frac{1}{2}\left( -\frac{q\Phi}{2\pi\hbar} \right)^2\frac{\hbar}{\m r^2}\delta t  - \frac{q\Phi}{2\pi\hbar} \frac{\hbar}{\m r^2} \delta t +\frac{1}{2}\frac{\m r^2}{\hbar}\frac{1}{\delta t}\epsilon_j^2  \ . \nn
\ee
We see that the linear terms in $\epsilon_j$ indeed vanish, demonstrating that we have identified the path which minimises the action. Moreover, since the action is quadratic note that this expression gives the total action.

Now let's use that the time interval $[0,t_R]$ has been divided in to $N$ steps of length $\delta t$ and that $t_R =  \frac{4\pi \m r^2}{\hbar}$ (note that $t_R$ goes as $1/\hbar$), to rewrite this action as
\be
\label{curvature}
A(\{k_j\}) & = & \hbar\sum_{j=1}^{N}\frac{1}{2}\left( -\frac{q\Phi}{2\pi\hbar} \right)^2\frac{4\pi}{t_R}\delta t  - \frac{q\Phi}{2\pi\hbar} \frac{4\pi}{t_R} \delta t +\frac{1}{2}\frac{t_R}{4\pi}\frac{1}{\delta t}\epsilon_j^2  
\\
& = & \hbar\frac{1}{2}\left( -\frac{q\Phi}{2\pi\hbar} \right)^24\pi  - \hbar\frac{q\Phi}{2\pi\hbar} 4\pi  + \hbar\frac{1}{8\pi} N\sum_{j=1}^{N}\epsilon_j^2 \ . \nn
\ee

Our aim is examine, in the classical limit $\hbar\to0$, how much a path can deviate from the path which minimise the action before its contribution to the amplitude is wiped out by destructive interference. 
The destructive interference in the path integral sets in when the deviation introduced by the $\epsilon_j$'s causes a change of $A/\hbar$ which becomes much greater than 1. Therefore as long as the sum over $\epsilon_j^2$ in the second line of (\ref{curvature}) is of order $1/N$, the corresponding path contributes to the constructive interference in the path integral.  If for instance $\epsilon_j=\epsilon$ for $1\leq j\leq N/2$ and $\epsilon_j=-\epsilon$ for $N/2+1\leq j\leq N$, then
\be 
& & N\sum_{j=1}^{N}\epsilon_j^2 = \nn  N^2 \epsilon^2 \ .
\ee
For this to be of order $1$ all we need $\epsilon\sim1/N$. This implies that a path can get a distance of order 1 away from the path which minimises the action, while the path still contributes constructively to the amplitude: First take $N/2$ increments of size $\epsilon$ and subsequently take $N/2$ increments of size $-\epsilon$ back to the path which minimise the action. This path deviates of order $\epsilon N/2$ from the path which minimise the action and since $\epsilon$ can be of order $1/N$ this deviation is of order 1. As the maximal magnitude of $\phi$ is $\pi$ this implies that paths which goes through all values of $\phi$ continue to contribute to the path integral in configuration space even in the classical limit $\hbar\to0$.

To summarise: One might have thought that since there is a physical quantum system which realises the QPE algorithm, then it should also have a classical limit where the path which minimise the action dominates the path integral. 
We see however that in the classical limit $\hbar\to0$ the return time $t_R\to\infty$ and that paths with all values of $\phi$ contributes.
\bigskip

\section{Summary and outlook}

As stressed in \cite{EHKM} computers (classical or quantum) are physical systems and computation is a physical process. Here we have identified a physical quantum system, the particle on a ring with a flux through, for which the physical process of time evolution, preforms the quantum computation of the quantum phase estimation algorithm. This system is not a universal quantum computer, but rather can be thought of as a dedicated quantum computer device which can perform quantum phase estimation. The first register is made up by the position (or after the quantum Fourier transform the angular momentum) states of the particle on the ring. To realise the quantum phase estimation algorithm for a $U(1)$ unitary operator it is sufficient to consider a standard electromagnetic flux. The full quantum phase estimation algorithm is realised with non-abelian flux: The unitary operator is encoded in non-abelian Aharanov-Bohm phase factor, and the controlled unitary operations are realised by the windings around the ring of the components of the particle with definite angular momentum.  

Having a physical system which realises the quantum phase estimation algorithm gives a more intuitive way to think of the algorithm and allows us to ask questions such as what happens in the classical limit $\hbar\to0$. To address this we have derived the path integral in configuration space for the quantum phase estimation algorithm as implemented by the Aharonov-Bohm effect for a particle on a ring. For this specific physical implementation, we have found that the time it takes to complete the algorithm diverges as $1/\hbar$ and that the path which minimises the action does not dominate in the classical limit. Thus there is no simplification to a single classical path in the classical limit. 

The novel link between the quantum phase estimation algorithm and the non-abelian Aharanov-Bohm effect established here potentially opens for a number of new ways to analyse the quantum phase estimation algorithm. We hope that further studies along these lines can help cast more light on the quantum phase estimation algorithm and help advance the field of quantum computation.

\bigskip

\acknowledgements
 We thank Klaus Mølmer, Peter K.~Jeppesen, Jan W.~Thomsen, Svend K.~Møller, Frederik R.~Klausen and Anna~L.P.~Bjerregaard for discussions. This work is supported by the Novo Nordisk Foundation, Grant number NNF22SA0081175, NNF Quantum Computing Programme. 
 
\appendix

%%%%%%%%%%%%%%%%%%%%%%%%%%%%%%%%%%%%%%%%%%%%%%%%%%%%%%%%%%%%%%%%%%%%%%%%%
%\newpage

\end{document}